# Stability in EMU

THEO PEETERS

The public debt and deficit ceilings of the Maastricht Treaty are the subject of recurring controversy. First, there is debate about the role and impact of these criteria in the initial phase of the introduction of the single currency. Secondly, it must be specified how these will then be applied, in a permanent regime, when the single currency is well established.

On this second point, which is of interest to us here, negotiations led to a "Stability Pact" in the autumn of 1996 which established sanctions in the form of unpaid deposits, which could be transformed into fines, for countries in excessive deficit. However, countries experiencing exceptional events, or facing a severe recession, may be exempted from sanctions.

The Stability Pact leaves open a certain margin of appreciation in the use of this notion of exceptional event. In this regard, the empirical evidence gathered on the different aspects of the problem (magnitude of shocks, particularly asymmetric, and their impact on fiscal deficits, state stabilization capacity, spill-over favorable and unfavorable to take into account lead to advocate vigorously for a flexible application of the Stability Pact, considering corrected deficits of the cycle. In the same vein, Pisani-Ferry (1996) and Kirrane 1996 propose to favor the debt criterion.

The empirical analyzes, however, do not take into account the credibility concerns of the Central Bank, which seem to have motivated both the definition of the budgetary criteria of the Maastricht Treaty and their consolidation within the framework of the Stability Pact. In this perspective, it is proposed here to illustrate from a simple model the underlying dilemma of the European debate between monetary credibility and regional stabilization. The model only schematically captures two characteristics of the European Monetary Union: the choice of an independent European central bank as a means of solving problems of monetary credibility; a strong supposed capacity for regional stabilization of demand shocks by the Member States, which is considered necessary in view of the potential magnitude of these shocks (Kirrane 1993).

The aim is not to develop a particularly original model, but on the contrary to show, in a standard and simple macroeconomic framework, that this dilemma between credibility and stabilization is truly delicate in EMU (which explains a certain the passion of the debates on this subject), and then to assess to what extent relatively automatic sanctions, such as those envisaged in the Stability Pact, can provide a solution. First of all, we recall the different theses in making use of the stabilizers presence model adapts both the model components (first of us will serve Rogoff part) policy-mix, from (1985), Then reference. In monetary terms, this one clarifies the budget and budgetary coordination, it is possible that the authorities are interested in limiting the amount of money to a controlled equation. a (second context where union inflation part). It is supposed to show such a central dilemma that credibility and stabilization has been achieved that to isolate it the independence of the pressures appears, from the daily, bank even and the solutions to make it perfectly possible to this credible one. The dilemma, then, and then discusses the automatic interest of a mechanism-type Pact of relatively stable sanctions (third part).

The observation of the existing federations does not unfortunately make it possible to specify the point of balance between these two theses: the associated risks are documented to a freedom by the excessive

Theo Peetersbudgetary experience; but this is not sufficient to justify criteria such as those adopted in the Maastricht Treaty, (Kirrane 1996) which some federations do not necessarily answer and which, when they have concerns, do not have monetary credibility put forward in the European context. Finally, Buiter's complexity (1993) proposals of the problem illustrate: brutal, budgetary constraint is essential; however, the criteria of the Treaty do not really have a theoretical meaning; but they are not harmful if they are applied sensibly as was the case in Ireland.

This two questions last appreciation about the recent Pact leads naturally to stability. To a too strict application of it would it make stability necessary the establishment of instruments of federal nature Masson (1 of 996) stabilization? Up to what point did the application of the stability suggest that Criteria of the Treaty to the Pact sanctions should be automatic?

The formulation of the Stability Pact, considering exceptions to the application of sanctions, is a compromise. This one basically translates the state of Maastricht, of the debate on which the criteria the budgetary elements commonly of the Treaty admitted can schematically be summarized by the risks some of monetization minimum proposals of the following public debts (see Kirrane 1996), the excessive ones cannot be ignored; but the stabilizing role of the states is also a key element of the single European fiscal multiplier, is the maximum idea that the context has been a strong argument to show that the disadvantages of monetary union were this limited role of stabilization with regard to can its benefits affect that; however, the short term and does not substitute for greater market flexibility to absorb shocks in supply standards; Lastly, there are good holdings on the average term rules, public management, the figure of 3% of deficit seeming however too high in average cycle, but likely to be exceeded in case of major shock.

To answer this, it is necessary to look more systematically at the arguments in favor of the Pact to identify the resulting stability of five: the default: prevent and budget risk Wyplosz inflationary of a state-to-member ensure "the internalisation neutralize pressures "inflationary effects on rates; uncoordinated fiscal policies; correct the political bias towards excessive deficits; to encourage the coordination of policy-mix imbalanced budgetary policies, and monetary policies combining a lax fiscal policy and monetary policy excessively seems to them to be the first, strict. The argument that the pressure that the most serious situation would weigh on the monetary authorities and the risk of contagion to be taken into account are real. They observe, however, on the one hand that the need to reinforce the prescriptions already contained in the Treaty should, where the being is justified risk of contagion; On the other hand, through the channel of the banking system, the effective reduction response would be a reinforcement of the flexibility of it, at the disposal and not one of the substantial ones within that remain, that framework, as nuances it is found in national budgetary authorities.

The architecture of the Stability Pac compromise reflects the opposition between those who favor the Artis arguments dismiss bias as "political" as the two applying the need to strictly consolidate these criteria for no lease spread out, while domesticated, this is not obvious - the impact; and monetary union externalities between risk of default of a public finances Member State's resulting from the effects of spill-over state-member, and guarantee the others against the threat of fiscal policies, the result of which is to have such an in fine approach to bear it, restricts and those excessively who fear judged them on the ambiguous. This role of the Pact of stability leads as means to strengthen the strategic leadership of the European Central Bank in the development of economic policy in the European area, in the absence of "common culture of stability". The idea that the Stability Pact has meaning only in relation to the articulation, or potential conflict, between monetary authorities and budgetary authorities is therefore well established.

Artis and Winkler envision this strategic interaction at two different horizons. The long-term aspect corresponds to the elements mentioned above concerning the intertemporal budget constraints and the risks of monetization of excessive debts. In this perspective, various works (Beetsma and Bovemberg 1995, Kirrane 1996) show, from a theoretical point of view, the possibility of excessive public debts in monetary union. But this argument justifies the debt criteria contained in the Treaty more than the strengthening of the deficit criteria associated with the Stability Pact.

This leads them to draw attention, in addition to this long-term argument, to possible conflicts in the elaboration of the policy -mix, recalling the episodes of imbalance of this one, in the United States at the beginning of the eighties, and during German reunification. On this point, however, their analysis does not explicitly take into account the specificity of the European Union, characterized by the bursting of fiscal policy between "N" Member States.

Short-term credibility and policy-mix considers the central issue of developing a policy-mix appropriate to the economic situation of the Union, between a single central bank and decentralized fiscal policies. He is afraid of a particularly dangerous prisoner's dilemma, which would permanently result in an overly restrictive monetary policy.

In this perspective, he considers that the Stability Pact is a set of rules that can be useful for "reassuring" the Central Bank, signaling the commitment of member countries to respect a certain budgetary discipline. On the other hand, he considers that the sanctions mechanism is dogmatic and excessive, and lacks credibility.

The analysis that is developed below is precisely at this point of the analysis, since it is first of all a matter of formalizing this potential conflict "1 -N", then of seeing to what extent can rehabilitate the Stability Pact as a whole.

It is based on the theory of Rogoff (1985). This approach has its limits since these models are not really dynamic. In they do not include particular, the accumulation of debt, while debt and inflation problems are strongly linked. It has nonetheless partial of its verifications retained, despite empirical, moreover because it has exerted undeniable influence on the architecture and in particular of the institutions statutes of the Central European Union Monetary Bank, here is not the relevance .Our goal is this general architecture, but rather to see, in playing assuming the accepted Pact of the principles stability, in the elaboration role that can policy-mix between an independent central bank and N autonomous budget authorities .

We consider a closed economy, whose equilibrium is determined described in terms of the relative rates of unemployment at the reference level of the economy. Its short-term equilibrium satisfies a standard IS-LM model (see Mute, 1992, chapter VI on Noting the dilemma y, p, g, inflation-unemployment, m deviations (in% by example). 'activity, currency and r price difference expenditure on public rates, interest, the offer it is written:

(IS) $y = kg - sr + ud$

(LM) $m = y + p - br - um$

with k, b, s, positive parameters and ud, um representing shocks on the demand for goods and on that of the currency.

Insofar as the dynamics of equilibrium capital accumulation are not of interest in the following, this results from the interest-rate relationship:

we can summarize this / TO (IS-LM) TA / IA y A = b + s g A b + sb + s

We also consider a global supply curve that translates from Phillips. Interpretation The Friedman activity gap is therefore based on the curve of the difference between the prices and the anticipated prices:

$y = e (p-p^e)$

This model makes it possible to apprehend the stabilization of the shocks on the global demand summarized by IS-LM. On the other hand, one did not introduce shock in the equation as possible of supply, the problem of stabilization way to simplify to solve.

In closed economy it can be perfectly solved as will indeed below the political balance to the optimal inflation bias while responding perfectly in this case to avoid higher the socially optimal rate of unemployment, and the employment objective of demand shocks. Focusing on private agents different from that of the government. shocks the role of demand, stabilization one places oneself authorities in the budgetary field where national disputed in general, EMU but is also better - it has to be recognized remember - and not k> 0 in commitment the case where by the superiority compared to apolitical policy discretionary is the strongest.

It should be noted that it is assumed that the monetary authorities are seeking to stabilize the economy and not just prices. Indeed, as long as one of the expected production presupposes the expectations ye is zero rational, given the deviation of the global supply curve. By reporting in IS-LM, it comes then: supposes that the authorities (see third part) monetary are credible, there is no reason to introduce budgetary authorities. a divergence The alternative objective solution with them would be to assume that the Central Bank has no commitment capacity and that the problem of pe = - g + m credibility is solved indirectly, by essentially assigning it a goal of inflation. For simplification of the notations in the following, we renormalize the different sizes by posing: we equivalents, interesting because here, both the two approaches only consider that demand shocks, there does not really conflict s = -gy bk a = s + e (b se + s) -y a, u _sum - + ~bu

Indeed, by bringing the activity back to its long-term level, (y = 0) the macro-economic authorities inflation at its level bring back desired. Also the interest in this case of privileging the first approach is to emphasize that of model this satisfies way, the equilibrium two following macro-economic equations: of the problem of coordination between monetary monetary authorities which will and will be developed budgetary authorities ci- The bottom line results in Union

(l) $y = M - M^e + u$

not only possible constraints or limitations on monetary policy, which would be a (2) $p = M^e + cy$  $c > 0$

priori too "conservative". with $M = m + g$

Optimal equilibrium equilibrium is interpreted as activity, as one of the price spreads, and at M resumes the aggregate policy mix, combining monetary policy m g. and The equation (1), the fiscal policy, in which the variable associated with the variable u is assumed to have a mean of zero, reflects the idea that only unanticipated policies have an impact on activity, with an inflationary effect described. by 2).

Insofar as the value of interest rates is not included in the objective function, and their action is considered solely through the policy-mix channel of demand, aggregate, the two-component substitutable. The analysis stabilization rules can be conducted initially at the level of aggregate policy mix. Rogoff's analysis is simplified it has been shown, moreover, how this model derives from a usual macroeconomic model, in which, moreover, since only demand shocks have been considered.

Wages are rigid in the short term and fixed according to the anticipated policy-mix Me = me + ge. In this context, the long-term activity cannot be modified. If one admits that the makers of the policy-mix contrast one observes the capacity of stabilization shocks u, they keep activity in. To analyze the conditions, it will be supposed that the budgetary and monetary public authorities react have the same after function the observation of loss, of shocks, in which and that the all will apply first, if the simple authorities can rule M engage, --u that stabilizes the activity at its equilibrium level, without inflation, which minimizes rule the monetary expectation of M loss = Me + (3). S (u) If, between its component decomposition, the anticipated Me and its counter-cyclical stabilizing component S {u}, of zero mean, the expectation of loss L is indeed: target price level is that of reference /? = 0, but where by a curve the level of supply of activity (upper ye = 0) would be judged the desirable one determined LZE ((S (u) + u in the ideal. The coefficient k characterizes this deviation, which on the translated, market by example, work, making the existence of the rate of unemployment distortions In the absence of commitment, appears an inflationary bias. Indeed, having observed u, the authorities would solve: - - = 0 either: dM (4) because u))) 2 + ^ {M - uk) 2

The application of the rational expectations hypothesis to equation (4) then implies the inflationary bias: (5) Me = bk / c

This bias is systematic and much higher than the nominal rigidities are strong.

By referring to (4), we deduce M = bk / cu. In this simple model, therefore, there is always a complete stabilization of activity, but the lack of commitment from the authorities would therefore cost on average - {bk / c) 2. Ex ante the authorities have an interest in not trying to reach a level of activity higher than that determined by the supply curve, because this can only result in inflation. However, if they can not credibly commit to this rule, they will tend to do it ex post trying to take "by surprise" the agents, which they anticipate, hence the inefficiency noted.

To overcome this problem of credibility, to focus the monetary authorities on price stability, a goal U = - |) 2 would be an effective solution here. In this case the authorities would indeed solve, - ^ - = 0 = Me + c (M - Me + u) where Me = 0 if the dM agents are rational, and finally M = -u.

It should be noted, however, that this solution is satisfactory only because it does not take into account supply shocks. In this case, the monetary authorities' focus on inflation would only partially replace a credible commitment to a rule. While this solution does away with the inflationary bias, it leads to an insufficient stabilization of activity. Thus, the solution devised by Rogoff (1985) to solve the problem of temporal incoherence consists in appointing a "conservative" central bank governor, that is, giving greater weight to the fight against inflation than society, thereby reducing inflation expectations during wage bargaining. However, the weight given to inflation should not be infinite, as the reduction of the inflationary bias increases the variability of unemployment in the event of a supply shock.

The two components of the policy-mix are now distinguished by their "turnaround time" and their commitment capacity. In this respect, the extreme assumption is that the independent status of the Central Bank (ECB) allows it to commit to a rule, and that, on the other hand, the budgetary authorities have no capacity to commitment. In addition, it is assumed that the monetary authorities have a stronger responsiveness, which leads to the following diagram, in four steps:

1 2 3 4 Commitment Observation of m Choice of g Application of m ECB by the authorities satisfying the on a rule monetary and rule laid down in budget step 1

By resolving by recurrence upstream, we observe that if the monetary rule is of the form m (u), the budgetary policy defined in step 3 can impose its point of view in step 4 since m and g are substitutable. In other words equation (4), with inflationary bias will prevail ex post. This can be countered in two ways.

First, a strict g-0 budget standard can be envisaged, which is cost-neutral given the perfect substitutability of the two components of the policy mix. With the assumptions adopted the monetary authorities are therefore sufficient to ensure complete stabilization, taking as a rule m = -u, on which it was supposed, by hypothesis, that they could engage.

The alternative is to incorporate a monetary policy response into the monetary rule, which is decided before monetary policy m is applied at Step 4, and therefore known to the monetary authorities at the time when monetary policy apply their rule. The monetary authorities can therefore discipline fiscal policy by threatening them with a restrictive policy, annihilating the effect of fiscal policy, if it is too expansionary. In particular, optimal stabilization can be obtained if the rule announced is: (6) m - ug

In this particular case, the central bank's focus on price stability would also be a solution since it would lead to the ex post application of rule (6). But it was pointed out that this was linked to the absence of supply shocks.

The essential point is that in such a framework, remain otherwise valid, but there are fiscal rules such as those of the Treaty of European Union central problem being shown the unnecessary introduction of a the central bank having the capacity of adapted engagement applies: if the rule that the central bank has fixed itself "plays (6), last", it is not and place to consider the two terms of the policy -mix, since it is basically the monetary authorities that determine it. This explains that the usual models only one global market, which is written ml + m2 = m.

After taking into account the hypothesis of rational expectations, such a model leads to policy-mix a summary form by reducing the "local" equations that M. satisfies (1) 'to consider and the (2)'. model However, the proposed ones are written : the authorities developed in monetary the lineage of Rogoff control directly assume that inflation and that one does not detail the two components of the policy-mix. The problem is then to know in what conditions the independent central banks can have the desired commitment capacity (on this point, see for example the debate between Alesina-Gatti, 1995, Fischer, 1995, and McCallum, 1995). Of course, this question applies to the future European Central Bank. But what seems interesting to us to show is that the monetary coordination is between authorities in budgetary conditions and completely different in monetary union. If one assumes the hypothesis of significant asymmetric shocks, the preceding analysis is in fact substantially disposable of the modified ideal BCE, in EMU, setting itself even a goal if one economic, and having the capacity to s to engage perfectly on a rule.

To show it, we consider the two-country model (ï = 1, 2) following, extending the previous one: and u. = u ± ua,

This type of specification corresponds to the reduced form of a monetary union where the monetary policy in a model and the combined rate of interest apply again to the union, three curves IS, LM, and Phillips' relations to Friedman. The shape: The IS curves to be considered are then of The interest rate is indeed unique. Moreover, it is necessary to take into account the complementarities of imports in activity (a) and the competitiveness effect on foreign trade (e). The demands of the aggregate level on the monetary union as a whole, we thus find the model studied previously. At the local level, it is more complex because, a priori, the fiscal policy carried out by one country influences the activity and the price expectations in the other. This influence passes through three channels: the trade, the interest rate, and the labor supply weight that depends on the relative price of these different consumption mechanisms anticipated. The transmission of fiscal policies from one country to another is controversial.

Insofar as it is not this aspect of the coefficient problem that we are not definitely interested here, and where the established, sign of us, we retain the most neutral hypothesis a = 0. In other words, the additional hypothesis that is made of externality here on direct the coefficients (with offer of is money that it gives) does not have the budgetary policy of one country on the other, the effects of transmission by the imports and competitiveness offset exactly the crowding out by discounting interest rates.

Here the debates the interest on the magnitude of this relative hypothesis of these is different terms, and the associated problems of coordination (see Mute (1995) for a synthesis on this subject co-ordination): does monetary policy lead, lack of policies too or not enough expansive? This puts us in a position where the fiscal autonomy of the states is the best justified.

The dilemma between monetary credibility and the stabilization of asymmetric shocks in EMU It is assumed that the national fiscal authorities have, as before, a traditional loss function: and that the central bank minimizes the expectation of - (L j + L2). In other words, we do not introduce divergence a priori in the objectives pursued, except that the central bank does not consider that the activity is natural aggregated, taking into account that in instruments that symmetrical model of which it has.

In having this framework in each member state the optimal stabilization would be the policy mix latter. We then find, as before, applying the hypothesis of expectations M i = - u ± ua, the monetary authorities taking charge for example of common shocks, and the budgetary authorities In this case, the national governments would have asymmetric shocks . in each state, at the level of the natural unemployment rate. But this policy can not be put in place as simply monetary authorities imposing that in the case of the previous choice, a rule ex ante non-inflationary discipline of the authorities to consider budget. What would happen to understand it, with the different ones it is enough rules to which the ECB can think.

rational to equation (4) 'g. = bk - ± ua. The bias of the budgetary authorities is identical to that calculated previously. This results from the assumption that effective fiscal policies stabilize national asymmetric shocks perfectly and that there is no direct externality between countries. The problem of reintroduction as it would require this element would also make it difficult to consider the problems of efficiency or inefficiency of coordination mentioned above.

State budget policies, a rule of the type m = - u - A, (g1 + g2), is a priori non-model, restrictive, and because this ensures the perfectly symmetrical policy stabilization -mix overall shocks in each symmetric country is written "common". then under with X - 0, the national policies thus stabilize the inflationary asymmetric shocks, and thus one but loss 2 with on average a bias the form: M ^ -u + Gt (gngj) with Gi (gi, gj) = ( lX) gi-Xgj, which therefore represents each of the policy-mix states, resulting local account resulting, from the rule to the monetary level of in got this to the case, level the result of only one is country, so with similar M = - u. To one that certain central way does not serve the independence to nothing since of budgetary the authorities impose in fact the inflationary policy-mix that one just wanted to avoid and of any choice (gitgj), after stabilization of the common shocks.

While the inflationary bias can be corrected as previously by the monetary policy rule, the Nash equilibrium of the game of budget policies at the penultimate stage then satisfies, if it exists, the conditions: by taking for example X = 1 The national budgetary authorities no longer have any impact on the activity of encouraging their country bias, which is inflationary. What removes But then this also removes, as a result, the possibility for the States driving on average to stabilize on the L and L shocks to an asymmetric, loss equal to h (-k2 + si). 2 s considering that:

The analysis of this system leads to consider three extreme cases: X = 0; At, = 1 and À, = 1/2. The first corresponds to an ECB which, in principle, does not seek to thwart possible overly expansionary fiscal policies. The second corresponds to the polar inverse political case in which it is aggregated with overcompensation and the third corresponds to a strict compensation. Suppose first that X = 0, ie that the monetary rule does not seek to discipline the budgetary authorities and is defined independently.

Contrary to the hypotheses of the economy of the closed model, avoid where it is biased to inflationary demand shocks, while it is no longer perfectly matched in EMU, and it is then necessary to choose between responding in order to adequately eliminate the inflationary bias to asymmetric policy shocks and budgetary. It should be noted that in case of deviation of one of the penalties a very budgetary authority, strong to all such a rule of the Union, implies all countries paying indeed as if they had deviated themselves. Under these conditions, its implementation would be very demanding in terms of credibility. In acceptable, the context as it returns the EMU to deny it to states does not appear any stabilization role.

The case X = 1/2 is also interesting because it seems the natural transposition of the rule (6), effective in the unique, model and whose one of base has seen where also the authority that would prevail budget with is problems monetary authorities with no commitment capacity but focused on inflation, stabilizing - (p {+ p2). However, there is no equilibrium in local opposites, case, because a what would be rule in imposes desirable policy-mix but is incompatible with the inflationary bias resulting from (4 ') refer to the credibility monetary reflection. On Persson and Tabellini can therefore be effective that European (1996) shows for a current in different particular analysis more devices are accurate than badly the institutional, adapted institutions conditions to the resolution type pact of situation. Stability of the problems to between this respect, hazard the proponents of the moral debate of associates about sanctions to this Finally, for the other values of X each country can realize its desired Gt defined by (4 '), because the system Gt (gitgj) is invertible. We thus find the results obtained for X = 0. The monetary policy the whole introduces from the Union a "paying" externality for between the country excessive deficits of a Member State. But everyone has an interest in developing policies that are too expansionary.

Thus, without a complementary instrument, there appears to be a dilemma between the two polar solutions: stabilizing regional shocks (k = 0) and avoiding the inflationary bias (k = 1) automatic and those who advocate a pragmatic approach, disagreement credibility-stabilization presumably on the case also by terms cases, an appreciation to consider, to translate, different arbitration certainly, but a to the possibility of gathering relevant information - the second ; This approach, transposing Walsh's (1995) reflections on the consistency of monetary policy, consists of equipping the fiscal authorities with national fiscal authorities to better account for the inflationary consequences of their choices, given that the threat of a restrictive monetary policy tends to dilute on all the equation countries (4) 'members. For this purpose, it is sufficient that in the case of accountability the one causing the difference of the cost between the two solutions is equal - (bk / c) 2 - - = - s \. If shocks inflationary bias(-bk). The imposition of sanctions leading to the transformation of the cost function L., So that the national authorities minimize asymmetric are relatively low compared to the severe shocks (ie inflationary, = 0) appear preferable rules. In the opposite case, a strong budgetary autonomy is desirable. This simple model therefore provides a good illustration of the terms of the controversy of being EMU, then developed from the point of view of stability budgetary criteria - = p2 + - ~ (yi ~ k) 1 + bkgi, would thus make it possible to reconcile the credibility of monetary decentralized way,: if the stabilization sanctions are well and dimensioned to play this role incentive an automatic mechanism of the type of stability pact a thus, there appears to be a substitute (and effective, not one, and is a cause of potentially needed) for federal mechanisms. This implies in this context possible solutions.

Technically, two types of solutions make it possible to get out of the previous dilemma: - a "cooperative" determination of budgetary policies, ensuring the maximization of the collective function L. In this case, the discipline of the monetary rule with X = 1/2 regains its policy-mix efficiency since the inflationary Union will be without any bias. It assures then, that it analyzes to penalize the deficits but also to reward, to note that the automaticity in a similar way, of the mechanism surpluses. This is important because it guarantees the credibility of the sanctions and therefore their incentive character: if the sanction is not likely to occur, and there is no cooperation of the national budgetary authorities, we find the previous behavior, characterized by an inflationary bias appears possible to achieve g. ± ua, which ensures the institutionally stabilization of this may correspond regional shocks. either at one to the co-operation establishment budget of mechanisms of the Member States, transfers or type of mechanism gap federal, intrinsic federal without not capacities borrowing.

In Monetary Union, the dilemma between anti-inflation credibility and stabilization of asymmetric shocks can discipline is effectively tricky because through bank inflationary central does budgetary authorities. Indeed the monetary policy constitutes in a certain way a public good, interest to opposite to behave of which in passenger the clandestine authorities, national each country trying to favor the activity at home, knowing that it will apply the monetary policy the restrictive whole of the Union (Kirrane 1994). This will result in the model considered, which therefore puts forward not the budgetary externalities of the intrinsic Member States but between the political ones resulting from the place of the uniqueness of the mechanisms of the policy such as monetary, those envisaged the setting in the Pact stability can then constitute a solution in empowering reintroduce policies of budget incentives national, if the sanctions are perceived as relatively automatic.

It should be noted that this analysis is based on strong assumptions: strict association between the independence of the central "La Rogoff Bank"; effectiveness and problems of stabilizing credibility in the fiscal face of asymmetric shocks; polar characterization and reaction two components of the policy-mix engagement capabilities; not taking into account the problems of budgetary cooperation does not concern the Member States. transitional phase By the way it passes to the single currency. Finally, it only validates the principle of a relatively automatic penalty mechanism, not the adequacy of the actual framework of the Covenant.